\documentstyle [prl,aps,epsf,twocolumn]{revtex}
\def\be{\begin{equation}}
\def\ee{\end{equation}}
\def\bea{\begin{eqnarray}}
\def\eea{\end{eqnarray}}

\def\nn{\nonumber\\}

\def\r#1{(\ref{#1})}

\newcommand{\p}{\partial}

\newcommand{\la}{\langle}
\newcommand{\ra}{\rangle}

\begin {document}

\draft
\title{Incommensurate spin correlations in Heisenberg spin-1/2 zig-zag ladders}
\author{Alexander A. Nersesyan$^1$, Alexander O. Gogolin$^2$ and
Fabian H. L. E\char'31ler$^3$}
\address{
$^1$Physikalisches Institut der Universit\"at Bonn, Nussallee 12,
53115 Bonn, Germany\\
$^2$Department of Mathematics, Imperial College, 180 Queen's Gate,
London SW7 2BZ, United Kingdom\\
$^3$Department of Physics, Theoretical Physics, Oxford 
University\\
1 Keble Road, Oxford, OX1 3NP, United Kingdom}
\date{Draft: \today}
\maketitle
\begin{abstract}
We develop a low-energy effective theory for spin-1/2 frustrated 
two-leg Heisenberg spin ladders. 
We obtain a new type of interchain coupling that breaks parity symmetry.
In the presence of an XXZ-type anisotropy, this interaction gives rise to 
a novel ground state, characterized by incommensurate correlations.
In the case of a single ladder, this state corresponds to a spin nematic
phase. For a frustrated quasi-one-dimensional 
system of infinitely many 
weakly coupled chains, this state develops true three dimensional spiral
order.  We apply our theory to recent neutron scattering experiments on
${\rm Cs_2CuCl_4}$.
\end{abstract}


Quantum spin chains have for a long time attracted the attention of both 
theorists and experimentalists. One of the main reasons for this
continuing fascination is the dominant role played by quantum fluctuations
in these systems, which lead to a rich variety of observed physical 
phenomena. In recent neutron scattering experiments \cite{col} on the quasi
one-dimensional frustrated Heisenberg antiferromagnet ${\rm Cs_2CuCl_4}$, 
an intriguing type of spiral order was observed. The very existence of 
such spiral order, which is an {\sl incommensurate} structure, is puzzling
from a theoretical point of view. 
According to the standard 
lore, 
three dimensional ordering of quasi-1D systems
results 
from
stabilization of the dominant spin correlations in the
underlying 1D constituents. Therefore, 
3D spiral order
would require strong incommensurate 1D spin correlations. This is in
contradiction 
with the known properties of simple antiferromagnetic Heisenberg 
chains and ladders, where the only known mechanism for generating 
incommensurabilities is via external magnetic fields.

In the present work 
we propose
a novel mechanism that naturally
gives rise to incommensurate correlations in spin ladders and quasi 1D 
materials in the absence of external fields.
Interestingly, we find that this phenomenon occurs in a standard model 
of two coupled spin-1/2 zig-zag Heisenberg chains, which has attracted
much recent interest \cite{aff,as}.

In the bulk of this letter we shall concentrate on this simple frustrated
ladder system in order to clearly exhibit the precise nature of the
mechanism 
of incommensurate spin correlations.
The application of our results to the quasi 1D case relevant for
${\rm Cs_2CuCl_4}$ is briefly discussed at the end.
In what follows we derive the low-energy effective theory for the
zig-zag ladder system, which we find to contain a parity-breaking
interchain interaction. This ``twist term'', which was not considered in 
previous studies, is an operator of conformal spin 1 \cite{alexei}
and
has important consequences: We will show that,
although it alters the position of phase boundaries 
already in the spin rotationally
invariant (SU(2)) case, it leads to incommensurate correlations only
in the presence of an easy plane XXZ anisotropy.

The Hamiltonian of the anisotropic zig-zag Heisenberg ladder is
\bea
{ H}&=& J\sum_{j=1,2}\sum_{n}
\{{\bf S}_{j,n}\cdot
{\bf S}_{j,n+1} +(\Delta-1) S^z_{j,n}
S^z_{j,n+1} \} \nn
&&+ J^\prime\sum_{n}
\left\lbrace
{\bf S}_{1,n}+
{\bf S}_{1,n+1}\right\rbrace
\cdot
{\bf S}_{2,n}\nn
&&+J^\prime\sum_{n}
(\Delta^\prime-1) \left\lbrace S^z_{1,n}+S^z_{1,n+1}\right\rbrace\cdot
S^z_{2,n}\,
\label{hamil}
\eea
where $|J^\prime|\ll J$ are the exchange couplings, and 
$|\Delta|\leq 1, |\Delta^\prime|\leq1$ are the anisotropy parameters.

In order to derive the low-energy effective field theory for \r{hamil}
we use standard bosonization techniques \cite{shel}.
In the continuum limit the spin operators decompose as
$
{\bf S}_{j,n}\to a_0 \left[
{\bf J}_j(x)+(-1)^n 
{\bf n}_j(x)\right]$, where
$a_0$ is the lattice spacing and $x=n a_0$. Here 
$
{\bf J}_j(x)$ and 
$
{\bf n}_j(x)$ are the smooth and staggered components of the
magnetization operator. In the framework of the bosonization method
these quantities are expressed in terms of canonical Bose fields
$\Phi_j(x)$ and their dual 
counterparts $\Theta_j(x)$, where
$[\Theta_j(t,x),\Phi_j(t,x^\prime)]=-i\theta(x-x^\prime)$.
Changing to symmetric/antisymmetric combinations $\Phi_\pm =
(\Phi_1\pm\Phi_2)/\sqrt{2}$, the Hamiltonian density consists
of several terms. The ``free'' part is 
\be
{\cal H}_0=\sum_{\sigma=\pm}\frac{v_\sigma}{2}
\left[(\partial_x\Theta_\sigma)^2+(\partial_x \Phi_\sigma)^2\right]
\label{free},
\ee
where $v_\sigma\propto J a_0$ are the spin velocities.
In addition, there is an in-chain current-current perturbation
\bea
{\cal H}_{JJ}&=&
g_1 \cos\sqrt{4\pi}\Phi_+\ \cos\sqrt{4\pi}\Phi_-\nn
&&+g_2 \sum_{\sigma=\pm}
\left[(\partial_x\Theta_\sigma)^2-(\partial_x\Phi_\sigma)^2 \right].
\eea
The interchain interaction gives rise to two perturbations which
we denote by ${\cal H}_{CC}$ and ${\cal H}_{PB}$ respectively
\bea
{\cal H}_{CC}&=&{g_3}
\cos\sqrt{4\pi}\Phi_+\ \cos\sqrt{4\pi}\Theta_-\nn
&&-{g_4}\sum_{\sigma=\pm}\sigma
\left[(\partial_x\Theta_\sigma)^2-(\partial_x\Phi_\sigma)^2 \right],
\eea
\bea
{\cal H}_{PB}&=&
g_5
\bigl(-\partial_x\Phi_-\sin\sqrt{4\pi}\Phi_+
+ \partial_x\Phi_+\sin\sqrt{4\pi}\Phi_-\bigr)\nn
&&+g_6
\partial_x\Theta_+\sin\sqrt{4\pi}\Theta_-\ .
\label{PB-bos}
\eea
The perturbation ${\cal H}_{CC}$ is the well-known current-current
interaction \cite{aff,as}, which promotes dimerization and leads to
the formation of a spectral gap. The perturbation ${\cal H}_{PB}$
is the novel parity breaking term, which in terms of the staggered
magnetizations 
$\vec{n}_{1,2}(x)$ 
reads
\bea
{\cal H}_{PB} &\sim& 
{\bf n}_1\cdot\partial_x
{\bf n}_2 -
{\bf n}_2\cdot\partial_x
{\bf n}_1 \nn
&+&(\Delta^\prime-1) \left[ n^z_1\partial_x{n}^z_2 -
{n}^z_2\partial_x{n}^z_1\right]. \label{tw1}
\eea
The in-chain coupling constants $g_{1,2}$ are determined by $J$ 
and $\Delta$, whereas $g_3 , \ldots , g_6$ are functions of 
$J^\prime,\Delta^\prime$.


Around the SU(2) symmetric point $\Delta=\Delta^\prime=1$, the 
bosonized Hamiltonian can be expressed in terms of four Majorana 
fermions \cite{shel,as}. The 
Hamiltonian of two decoupled chains simply becomes
$
{\cal H}_0 = -i\frac{v_s}{2}\sum_{j=0}^3
\left(\xi_R^j\partial_x\xi_R^j - \xi_L^j\partial_x\xi_L^j\right).
$
The perturbation $
H' = {\cal H}_{JJ}+{\cal H}_{CC}+{\cal
H}_{PB}$ is of the form
\be
H' = \sum_{j=1}^3\alpha_j\  A_j+
\sum_{j=1}^4\beta_j\ B_j,
\ee
where the operators $A_j$ and $B_j$ are given by
\bea
A_1&=&\xi^0_R\xi^1_L\xi^2_L\xi^3_L +(R\to L),\nn
A_2&=&\xi^0_R\xi^1_R\xi^2_R\xi^3_L+(R\to L),\nn
A_3&=&\xi^0_R ( \xi^1_R\xi^2_L+\xi^1_L\xi^2_R)\xi^3_R+
(R\to L),\nn
B_1&=&\xi^0_R\xi^0_L(\xi^1_R\xi^1_L+\xi^2_R\xi^2_L),\nn
B_2&=&\xi^0_R\xi^0_L \xi^3_R\xi^3_L,\nn
B_3&=&\xi^1_R\xi^1_L \xi^2_R\xi^2_L,\nn
B_4&=&(\xi^1_R\xi^1_L+\xi^2_R\xi^2_L)\xi^3_R\xi^3_L\ .
\eea
We note that in the above formulas we have neglected
terms that lead to a renormalization of the spin velocities
\cite{as}.
The couplings $\alpha_j,\beta_k$ are easily expressed in terms of the
$g_i$.
The operators $A_j$ originate 
from ${\cal H}_{PB}$, whereas the
$B_k$'s stem from ${\cal H}_{CC}$ and ${\cal H}_{JJ}$. 

\begin{figure}[ht]
\begin{center}
\noindent
\epsfxsize=0.45\textwidth
\epsfbox{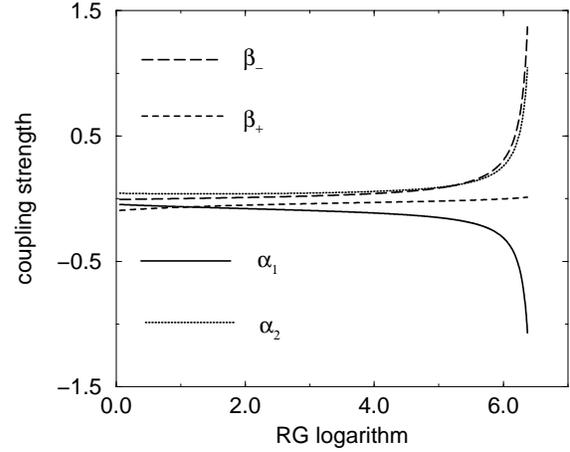}
\end{center}
\caption{\label{fig:su2}%
RG flow in the SU(2) symmetric case. There are only 2 independent
couplings for twist ($\alpha_{1,2}$) and current-current operators
($\beta_{\pm}=(\beta_3\pm\beta_1)/2$) respectively. The initial conditions
are chosen such that the couplings $\beta_\pm(0)<0$ and in the absence
of the twist operators would flow to zero coupling.
} 
\end{figure}

\begin{figure}[ht]
\begin{center}
\noindent
\epsfxsize=0.45\textwidth
\epsfbox{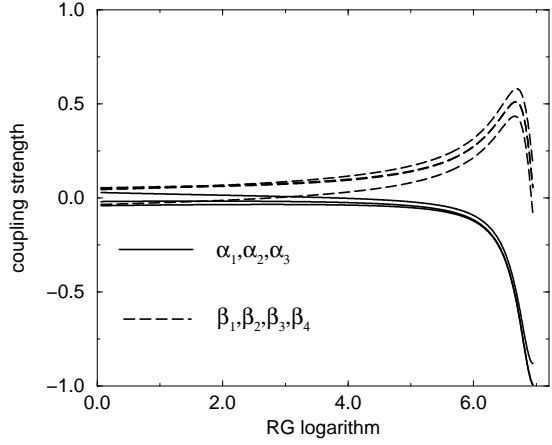}
\end{center}
\caption{\label{fig:twist}%
Example of an RG flow where the twist couplings reach strong coupling
first. The structure of equations \r{RGanis} is such that the twist
couplings cannot grow without 
``help'' of the current-current
interaction. 
} 
\end{figure}

All operators are marginal but while the $B_k$ have conformal spin $0$, the
$A_j$
have conformal spin $1$. We derived the renormalization-group 
flow for the perturbation ${\cal H}_{pert}$ from the Operator Product
Expansion for the perturbing operators in the spirit of
\cite{polyakov,as}
\be
\begin{array}{lll}
\dot{\alpha}_1 & = & 2 \alpha_2\beta_2+4\alpha_3\beta_1\\
\dot{\alpha}_2 & = & 2 \alpha_1\beta_2+4\alpha_3\beta_4\\
\dot{\alpha}_3 & = & 2 \alpha_1\beta_1+
2\alpha_2\beta_4+2\alpha_3\beta_3\\
\dot{\beta}_1 & = & -4\alpha_2\alpha_3 + 2\beta_1\beta_3
+2\beta_2\beta_4\\
\dot{\beta}_2 & = & -4\alpha_3^2+4\beta_1\beta_4\\
\dot{\beta}_3 & = & -4\alpha_1\alpha_2+2\beta_1^2+2\beta_4^2\\
\dot{\beta}_4 & = & -4\alpha_1\alpha_3+2\beta_1\beta_2+
2\beta_3\beta_4
\end{array}
\label{RGanis}
\ee
Here a dot denotes the derivative with respect to the RG logarithm
${\rm const}\times\ln(\Lambda v_s/|\omega|)$, where $\Lambda$ is a 
momentum cutoff\cite{foot1}.
The RG flow determined by \r{RGanis} is clearly complicated. Therefore 
we performed a numerical analysis of \r{RGanis}. Our findings are the
following: 

$\bullet$\ At the SU(2) symmetric point, the current-current interactions
reach the strong coupling regime first (in the cases where
the flow is towards strong coupling). This corresponds to
the dimerized phase found in \cite{as}. A new feature 
caused by the twist operators is that the tendency towards
dimerization extends into part of the
region of ferromagnetic current-current
interchain interactions (see Fig.~\ref{fig:su2})
$\bullet$\ Away from the SU(2) symmetric point {\sl both}
twist and current-current couplings generally (but not always) flow
towards strong coupling. However, it is now possible for the twist
terms to reach the strong coupling regime first (see
Fig.~\ref{fig:twist})! We take this as
evidence for the existence of a novel phase, the physics of which is
determined by the twist terms.

In order to elucidate the physics of this new phase we now turn to 
a mean-field analysis of the perturbations. 
The situation is particularly simple 
in the limit of very 
strong anisotropy,
$|\Delta|,|\Delta^\prime|\approx 0$, corresponding to two coupled XX 
chains. We therefore shall discuss this case first and only then 
generalize to arbitrary anisotropy.
Using the bosonization formulas for the XX point \cite{alexei,GNT} and
retaining only the relevant  part of the perturbing operator 
we arrive at 
\be
{\cal H}= {\cal H}_0 + \gamma \partial_x\Theta_+\sin\sqrt{2\pi}\Theta_-,
\label{xx}
\ee
where ${\cal H}_0$ is given by \r{free}.

In order to analyze \r{xx} we use a self-consistent mean-field
approach. Assuming that the ground state of the system is
found in the sector with a nonzero topological spin current $\p_x
\Theta_+$, we arrive at the following mean-field Hamiltonian
\be
{\cal H}_{MF}  = {\cal H}_0+\kappa \p_x \Theta_+ -
\mu \Lambda \sin \sqrt{2 \pi} \Theta_- ,
\ee
where
\be
\kappa = \gamma \la  \sin \sqrt{2 \pi} \Theta_-  \ra,~
\mu \Lambda =  - \gamma \la  \p_x \Theta_+  \ra .\label{se}
\ee
Thus ${\cal H}_{MF}$ decomposes into 2 commuting parts, 
${\cal H}_++{\cal H}_-$, with
\bea
{\cal H}_-&=&\frac{v_s}{2}\left[(\partial_x\Phi_-)^2
+(\partial_x\Theta_-)^2 \right] - \mu \lambda \sin\sqrt{2\pi}\Theta_-
\nn
{\cal H}_+&=&\frac{v_s}{2}\left[(\partial_x\Phi_+)^2
+(\partial_x\Theta_+)^2+\frac{2\kappa}{v_s}\partial_x\Theta_+\right].
\eea
The "$+$" channel is solved by eliminating the $ \p_x \Theta_+$ term
through a field redefinition: $\Theta_+(x)\to \Theta_+(x)-\kappa
x/v_s$. The average value of $ \p_x \Theta_+$ is then given by 
$
 \la \p_x \Theta_+ \ra = - {\kappa}/{v_s}. 
$
The "$-$" channel is a Sine-Gordon model for the dual field and can be
solved exactly, the expectation value of the mass term being
$
\langle \sin\sqrt{2\pi}\Theta_-\rangle = c(|\mu|/\Lambda v_s)^{1/3}
{\rm sgn}~\mu,
$
where the constant $c$ can be calculated \cite{luk}.
The self-consistency conditions \r{se} then lead to the solution 
\be
\mu=  \pm \Lambda v_s z^3 _0 c^{3/2}\ ,
\quad \kappa= \pm \Lambda v_s z^2 _0 c^{3/2}\ .
\label{mu-kappa}
\ee
where $z_0 = \gamma / \Lambda v_s$ is a dimensionless coupling constant.
The expectation value of $\Theta_-$ is determined by the position of
the minima of the Sine-Gordon potential; 
$
\langle \Theta_-\rangle = \sqrt{\pi/8}\ {\rm sgn}~\mu ({\rm mod}\ \sqrt{2\pi})\ 
.
$
This allows us to express the dual fields in chains 1 and 2 as
\be
\Theta_{1,2}(x) = \frac{1}{\sqrt{2}}\Theta^0_+(x)-\frac{\kappa
x}{\sqrt{2}v_s} \pm \frac{\sqrt{\pi}}{4}\ ,
\ee
where $\Theta^0_+(x)$ as well as its derivative have zero expectation
value. Using this together with the bosonization table at the XX point,
we find the following asymptotic behaviour for spin-spin correlation
functions in the strong-coupling phase
\be
\langle S_1^+(x) S_j^-(0)\rangle\sim \frac{(-1)^{x/a_0}}{|x|^{1/4}}\
\exp\left[-i\sqrt{\pi/2}(\kappa x/v_s)\right]
\label{asymp}
\ee
where $j=1,2$. The transverse spin correlations still fall off with a
power law but their decay is much slower than for an isolated XX chain
where $\langle 
n^+(x) 
n^-(0)\rangle\sim |x|^{-1/2}$.
Furthermore the correlations are {\sl incommensurate}! The
characteristic momentum of the magnetic spiral is
\be
q_0 = \pi/a_0 - \sqrt{\pi/2}(\kappa/v_s)\ .
\ee
The deviation from the anitiferromagnetic wave vector $\pi/a_0$ is
very small for a weak interchain coupling, which is in qualitative
agreement with \cite{col}.
We have performed a stability analysis of the above mean-field 
solution and found only convergent corrections, 
e.g. to the 
exponent $1/4$ in \r{asymp}.


The solution of the XX zig-zad ladder \r{xx} we obtained actually describes a
{\sl spin nematic} ground state of the model. This phase,
with unbroken
time reversal symmetry\cite{andreev}  is characterized
by nonzero local spin currents polarized along the anisotropy (z-)axis. 
The longitudinal (in-chain) component of the total spin current is given by
$
J^z _{\parallel} (x) = - \sqrt{2/\pi} v_s \partial_x \Theta_+ (x).
$
Using equations of motion for the spin densities, one easily finds the
transverse (interchain) part of the current, originating from the twist
term in \r{xx}:
$
J^z _{\perp} (x) = - \sqrt{2/\pi} \gamma \sin \sqrt{2 \pi} \Theta_- . 
$
>From the mean-field analysis above one immediately
obtains the important result that
$
\la J^z _{\parallel} \ra = - \la J^z _{\perp} \ra = \pm \sqrt{2/\pi} \gamma .   
\label{currents}
$
Thus, in the ground state the longitudinal and transverse spin currents are
equal in magnitude but propagate in opposite directions. The resulting picture
 shown in Fig. 3 demonstrates local currents circulating around the triangular
plaquettes in an alternating way, with  the {\it total} spin current
of the system being zero.

The spin nematic phase preserves the spin U(1) symmetry but spontaneously
breaks a special discrete $Z_2$ symmetry of the model. Indeed, the twist term
(\ref{tw1}) is invariant under a tensor product of the site-parity 
and link-parity 
transformations\cite{eggert} on the two chains,
${\cal P}_{12} = P^S _1 \otimes P^L _2$
[though it breaks 
$P^{S(L)} _1 \otimes P^{S(L)} _2$].
In the XX case \r{xx}, this transformation
reverses the signs of all currents, reflecting
the $Z_2$ degeneracy of the ground state. The spectrum of the system
is rich and, apart from gapless excitations in the "+" channel,
 contains massive quantum solitons and their bound states (breathers),
as well as kinks carrying fractional topological charge.


\begin{figure}[ht]
\begin{center}
\noindent
\epsfxsize=0.45\textwidth
\epsfbox{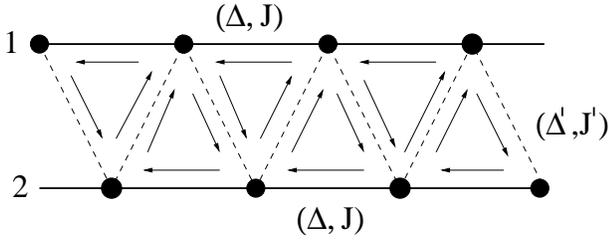}
\end{center}
\caption{\label{fig:current}%
Structure of the spin currents in the spin nematic phase.} 
\end{figure}

Moving in parameter space from $\Delta,\Delta^\prime=0$
(XX point) to $\Delta,\Delta^\prime=1$ (XXX point) in the general XXZ
case, we found the following: 

$\bullet$\ The spin-nematic phase 
occupies the whole region $g_4 \leq g_2$ beyond which
a gap appears in the $+$ channel as well, and
the spin correlations become short-ranged (although they remain
incommensurate). 

$\bullet$\ In the fully gapped phase, the spin currents and the
incommensurate wave vector $q_0$ are reduced and continuously vanish
as the SU(2) symmetric point is approached. Indeed, the spin nematic
state cannot be stable at the XXX point because nonzero spin currents
would break the SU(2) spin rotational symmetry, which is forbidden in
1 + 1 dimensions.

In light of the experimental results \cite{col} it is important to
investigate the effects of an external magnetic field. At the XX point
we find that a longitudinal field has essentially only 
trivial effects, whereas a transverse field leads to the formation of
a gap in the $+$ channel and completely destroys the spiral phase (above a
critical field). Indeed, the magnetic field oriented along the $x$-axis in
spin space enters the Hamiltonian
with the term\cite{kus}
\[
{\cal H}_{\em mag}= h \sum\limits_{j=1,2} \sin\sqrt{4\pi}\Phi_j
\cos\sqrt{\pi}\Theta_j
\]
where $h$ is proportional to the applied field.
This perturbation again has conformal spin $1$ and, together with
the twist term \r{xx}, leads to a complicated RG flow, where 
an effective XY anisotropy, described by
the following conformal spin zero operator, is generated\cite{kus}: 
$
{\cal O}_{\em XY} \sim h^2 \cos\sqrt{2\pi}\Theta_+\cos\sqrt{2\pi}\Theta_- .
$
This strongly relevant operator  locks the fields $\Theta_{\pm}$
at the vacuum values $\Theta_+ = \sqrt{\pi/2}$, $\Theta_- = 0$, or vice versa,
and gives rise to a commensurate Neel ordering of the spins along the $y$-axis, 
characterized by a nonzero average staggered magnetization
$\la n^y _1 + n^y _2 \ra \sim \la \sin \sqrt{\pi/2} \Theta_+ \ra \la
\cos \sqrt{\pi/2} \Theta_- \ra$.

Let us now turn to the case relevant for the experiments \cite{col},
where we have two-dimensional arrays of weakly coupled chains. 
We illustrate our findings for the simpler XX case. The
perturbation to the free part of the multi-chain Hamiltonian is
\bea
{\cal H}' =
\gamma\sum_j
\partial_x\left(\Theta_j+\Theta_{j+1}\right) \sin\left[
\sqrt{\pi}(\Theta_j-\Theta_{j+1})\right],
\eea
We perform a mean-field analysis in a way similar to the
two-chain case above. Assuming that 
the following averages
$\gamma \langle \sin\sqrt{\pi}(\Theta_j - \Theta_{j+1})\rangle =
\kappa$, $\gamma\langle\partial_x(\Theta_j +
\Theta_{j+1})\rangle=2\mu$ 
are nonzero
and redefining the fields as
$
\Theta_j = -2\kappa x + j \sqrt{\pi}/2 + \bar{\Theta}_j\ ,\
$
with
$
\langle \bar{\Theta}_j\rangle=0\,
$
we obtain 
a self-consistent 
mean-field Hamiltonian
\be
{\cal H}_{MF}={\cal H}_0[\bar{\Theta}_j] - 2\mu\sum_j
\cos\sqrt{\pi}(\bar{\Theta}_j - \bar{\Theta}_{j+1})\ .
\label{quasi}
\ee
The Hamiltonian \r{quasi} can be viewed as describing coupled
Josephson-junction arrays 
and leads to the pinning of
the fields $\bar{\Theta}_j$. The 
resulting average staggered
magnetization is given by 
\be
\langle n^\pm(x)\rangle = \exp(\pm i\pi j/2\pm 2\sqrt{\pi}\kappa x)\ .
\ee
This corresponds to 
incommensurate spiral order along the chains with
a $90^o$ rotation of the average staggered magnetization in the
transverse direction. This is in qualitative agreement with experiment
\cite{col}. Details of the above calculations as well as a
more quantitative comparison to experiment will be presented in a
separate publication \cite{unp}.

We thank P. Azaria, R. Coldea, R. Cowley, T. Giamarchi, P. Lecheminant
and A.M. Tsvelik for important discussions.


\begin{thebibliography}{10}
\bibitem{foot1}
We note that the presence of the operators with nonzero conformal spin
leads to a renormalization of the velocities at one loop order
\protect{\cite{unp}}.
\end{thebibliography}
\end{document}